# Anomalous Self-Generated Electrostatic Fields in Nanosecond Laser-Plasma Interaction

L. Lancia[1,2], M. Grech[3], S. Weber[1,4], J.-R. Marquès[1], L. Romagnani[1], M. Nakatsutsumi[1,5], P. Antici[1,2,*], A. Bellue[4,6], N. Bourgeois[1], J.-L. Feugeas[4], T. Grismayer[7], T. Lin[8], Ph. Nicolaï[4], B. Nkonga[6], P. Audebert[1], R. Kodama[5], V. T. Tikhonchuk[4] and J. Fuchs[1,†]

[1]*LULI, École Polytechnique, CNRS, CEA, UPMC, route de Saclay, 91128 Palaiseau, France*

[2]*Dipartimento di Scienze di Base e Applicate per l'Ingegneria, "Sapienza" Università di Roma, Via A.Scarpa 14-16, 00161 Roma, Italy*

[3]*Max Plank Institute for the Physics of Complex Systems, D-01187 Dresden, Germany*

[4]*Centre Lasers Intenses et Applications, CNRS - Université Bordeaux 1 - CEA, 33405 Talence, France*

[5]*Graduate School of Engineering, Osaka University, 2-1 Yamada-oka, Suita, Osaka 565-0871, Japan*

[6]*Institut de Mathématiques de Bordeaux, CNRS - Université Bordeaux 1 - Université Bordeaux 2, 33405 Talence, France*

[7]*Department of Physics and Astronomy, University of California, Los Angeles, California 90095, USA*

[8]*Fox Chase Cancer Center, Philadelphia, USA*

Electrostatic (E) fields associated with the interaction of a well-controlled, high-power, nanosecond laser pulse with an underdense plasma are diagnosed by proton radiography. Using a current 3D wave propagation code equipped with nonlinear and nonlocal hydrodynamics, we can model the measured E-fields that are driven by the laser ponderomotive force in the region where the laser undergoes filamentation. However, strong fields of up to 110 MV/m measured in the first millimeter of propagation cannot be reproduced in the simulations. This could point to the presence of unexpected strong thermal electron pressure gradients possibly linked to ion acoustic turbulence, thus emphasizing the need for the development of full kinetic collisional simulations in order to properly model laser-plasma interaction in these strongly nonlinear conditions.

[†]julien.fuchs@polytechnique.fr

Understanding the transport of intense laser beams through long underdense plasmas is crucial for Inertial Confinement Fusion (ICF) [1] and laboratory astrophysics [2]. Laser Plasma Interaction (LPI) mechanisms [3] that govern transport can be, depending on conditions, detrimental [4,5] or beneficial [6,7,8]. Among those are laser energy deposition in the plasma and plasma density perturbation induced by laser heating and/or ponderomotive force [3].

In this Letter, we take advantage of recent progress in electromagnetic field mapping using point-like ultrashort proton sources [9,10] to access, for the first time, the complete spatial distribution of plasma density and temperature resulting from the interaction of a long, high-power laser pulse with an underdense plasma. This is possible in a single shot and with high spatial and temporal resolution since the protons are sensitive to the electric (E) fields created by the electron pressure and laser intensity (ponderomotive force) gradients. This new technique stands in contrast with optical diagnostics [11,12] which cannot address, in detail, simultaneously the LPI inside as well as far away from the propagation region. Compared to expected ICF conditions [1], we work in a regime of lower temperature that is less non-linear and less kinetic. Nevertheless, in our experiment, in the regions of the plasma with strong temperature gradients, we observe anomalously strong E-fields around the laser pulse. These fields cannot be reproduced by state-of-the-art three-dimensional (3D) laser propagation code with non-linear hydrodynamic and non-local heat transport models. However, in the case where the temperature gradients are negligible, the E-fields (solely due to the ponderomotive force in this case) are well reproduced by the same code, which provides the first direct mapping of the ponderomotive force. With this new vision offered by proton radiography, our field measurements underline the need for further theoretical developments and inadequacy of present macroscopic simulation models to properly model LPI.

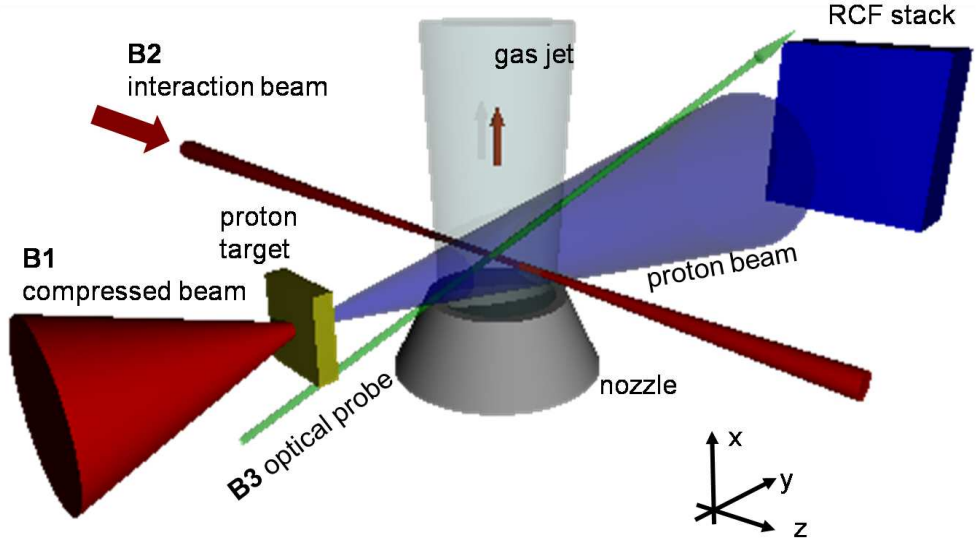

Figure 1: (color online) Setup of the experiment.

The experiment was performed using the LULI 100 TW laser facility. The laser beam, of $\lambda_0$= 1.053 µm wavelength, was split after wavefront correction [13] into three beams: B1 (350 fs) used for the probing protons generation [9]; B2, the interaction beam (10 J, 400 ps full-width at half-maximum, FWHM) used for plasma interaction, and B3, a low-energy probe for plasma interferometry. All beams were horizontally polarized. As shown in Fig.1, B1 and B2, delayed with sub-ps precision, were focused at 90° with respect to each other. B2 was focused onto a supersonic He jet using an f/24 lens with f=2.1 m, producing a Gaussian intensity distribution at focus $I_L = I_o \exp[-(r/r_L)^2]$ (where $r_L \approx 36$ µm and $I_o = 4 \times 10^{14}$ W/cm$^2$), with Rayleigh length of ~ 4 mm, which is larger than the interaction length in the jet (~2 mm). As observed by interferometry, the plasma was fully ionized around 200 ps before the peak of B2, and over a wide area (600 µm radially), with an electron density of $n_e$ ~ 0.016 $n_c$ ($n_c$[cm$^{-3}$]~$10^{21}/\lambda_0^2$[µm]) for 10 bar backing pressure (we took interferometric data, however, they are not shown as they just yield the global plasma density, but not the fine laser propagation details revealed by proton probing). Hydrodynamic simulations, performed with the code CHIC [14] using the experimental conditions, predict electron ($T_e$) and ion ($T_i$) temperatures of about 200 eV and 50 eV, respectively. The probing protons were generated at a distance d = 3.5 mm from the center of the gas jet and detected downstream at a distance D = 42.5 mm from the gas jet by a stack of films (RCFs [15]) that allows resolution in proton energy (determined using SRIM [16]). The proton radiography spatial resolution was ~5 µm [9], and its temporal resolution, given by the time the protons take to cross the width of the detected fields and the range of proton energies detected in a given RCF layer, was ~ 4 ps for 3 MeV protons.

A typical example of a proton radiograph obtained 120 ps after the temporal peak of B2 is shown in Fig.2(a). Fig.2(b-d) show line-outs, along the transverse *x*-direction (see Fig.1), of the 2D proton deflection map shown in Fig.2(a) at three different locations along *z*. We stress here that the map and the lineouts correspond to deflections patterns integrated through the plasma in the y-direction and projected onto the film. At the laser entrance in the plasma, one observes a single filament (Fig.2(b)) where the slowly varying relative modulation of the proton dose ($\Delta N_p/N_p$) has a maximal variation of 25% over ~50 µm. Deeper in the plasma, after propagation over ~1 mm, the global modulation is reduced to ~15% and small-amplitude modulations appear over ~ 10 µm (Fig.2(c)). This is a manifestation of the onset of filamentation [3], which appears fully developed over a distance of 2 mm, i.e. at the foot of the plasma density profile produced by the gas jet, as shown in Figure 2(d), with small-scale $\Delta N_p/N_p$ up to ~20%. The fact that there is a zone in the plasma, located at the edge of the gas jet (between z~1.4 and 1.6 mm), where there are strong density gradients and where the deflection pattern seems to be washed-out could be linked to a combination of effects. The small scale of densely grouped filaments is such that they are difficult to be resolved spatially; their unstationnarity also makes them difficult to be resolved temporally over the integration time of the diagnostic (~4 ps); and plasma density perturbations at the gas/vacuum interface can also reduce the electrostatic field (the density then evolves on the same scale as the laser-induced density perturbation). In short, we see here that proton radiography provides us with unique insights on the laser-plasma longitudinal dynamics on a single shot.

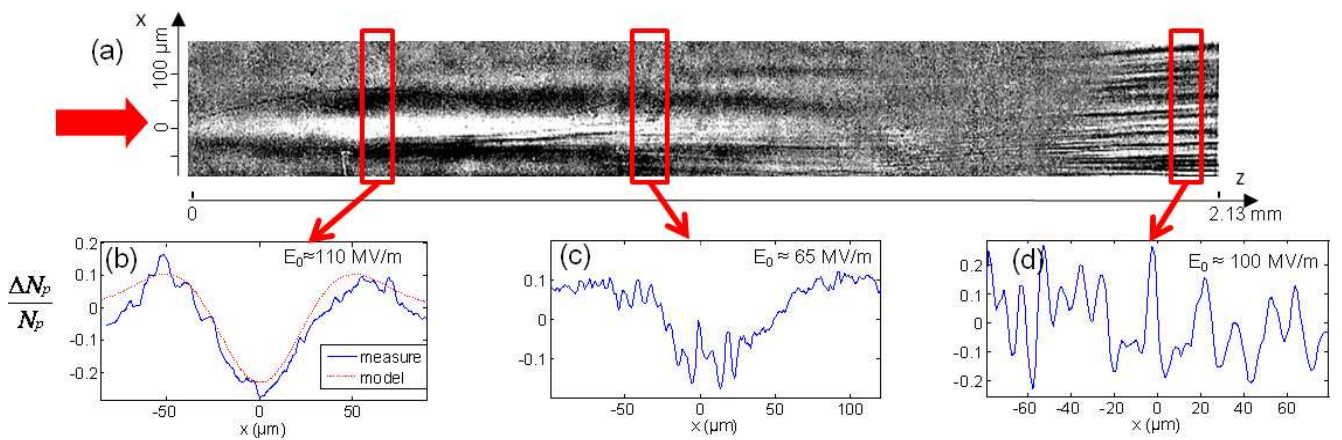

Figure 2 (Color online) (a) 3 MeV proton radiograph of the propagation of a laser filament 120 ps after the peak of the pulse in a 0.016 $n_c$ density plasma. (b,c,d) Transverse (*x*) lineouts of the proton modulations extracted at various positions. Scales are related to the gas jet plane.

Due to the low plasma density, scattering effects on the proton trajectories are negligible and proton deflections are therefore due to self-generated, quasi-static, E and/or B fields. Due to the experimental geometry, only the $E_x$ and $B_z$ components could be measured. In our conditions, most of the mechanisms will lead to quadrupolar (or even higher order) $B_z$ magnetostatic fields [17], but quadrupolar or even azimuthal cannot yield proton deflection patterns reproducing the observation (at least in the unfilamented region), as attested by particle-tracing simulations. Moreover, fields of the order of tens of Tesla with an energy density $\sim 10^{-1}$ J/mm$^3$, i.e. higher than the thermal one, would be necessary to observe modulations in the proton signal (however of different pattern). The proton radiography is therefore sensitive, in this experiment, only to $E_x$.

In our conditions of a quasi-static, quasi-neutral plasma, the equation of motion along x for the electron fluid, neglecting the electron inertia, writes: $\sum F = (1/n_e) \partial P_e/\partial x + 1/[2cn_c] \partial I_L/\partial x - eE_x = 0$ (with $P_e = n_e T_e$), as the thermoelectric term [18] is here negligible. Hence, the required balance of the electron fluid under the action of the laser and thermal forces results in a restoring E-field that can be separated in electron pressure and laser ponderomotive components $E_x = E_{ep} + E_{lp}$. It is important to note that the pressure is composed of two terms of different physical origins, a temperature gradient component ($\nabla T_e$) and a density gradient component ($\nabla n_e$) that can act in opposite ways since in general the density has a dip on axis while the temperature peaks on axis. Evaluating the effect of the pressure therefore requires precise determination of both components. Nevertheless, in general, as $I_L$, $n_e$ and $T_e$ can all be assumed to be Gaussian shaped, therefore, $E_x$ should take the form of the x-derivative of a Gaussian: $E_x(x,y) = (2e_N)^{1/2} E_0 (x/R) \exp(-(x^2+y^2)/R^2)$, with $e_N = 2.71$.

To extract information about the magnitude of the E-field from the radiograph, we need to relate $E_x$ to the observed proton signal projected on the RCF detector. For y-propagating probe protons having energy $\mathcal{E}_p$, the angular deflection writes $\theta_d(x) = e \int dy\, E_x/(2\mathcal{E}_p)$, integrating over the proton trajectory. Hence $\Delta N_p/N_p = -(D/M)(d\theta_d(x)/dx) \sim -eD/(2M\mathcal{E}_p) \int dy\, dE_x/dx$, with $M = 1+D/d \sim 13$ as the proton projection magnification on the RCFs.

The assumed analytical form of $E_x$ corresponds well to the observed $\Delta N_p/N_p$ in the first 1 mm of the propagation (see the dotted line in Fig.2 (b)). From Fig.2(b), we can infer that R~40 µm which is consistent with the laser beam transverse size $r_L$ and with $|\Delta N_p/N_p|^{max} = (2\pi e_N)^{1/2} D e E_0/(2M\mathcal{E}_p)$ we estimate that $E_0^{start}$~110 MV/m. In Fig.2(c) the small scale modulations related to the onset of filamentation, similarly, correspond to E-field of the order of a few MV/m. Once the filamentation is fully developed (Fig.2(d)), the spatial scale of the fields is 10 µm while their amplitude remains of the order of few tens of MV/m, and up to $E_0^{end}$~100 MV/m.

We now compare these results to numerical simulations of LPI performed using the 3D code PARAX [19]. This code treats wave propagation in the paraxial approximation. It is coupled to a nonlinear, two-temperature ($T_e$,$T_i$), single fluid hydrodynamic module, which is two dimensional in the transverse $x$,$y$ plane [20]. The simulated plasma volume is 2 mm in $z$ and 256 µm in $x$ and $y$, and experimental plasma density as well as laser intensity profiles are used together with initial temperatures as simulated using CHIC. We stress that both the experimentally measured transverse and longitudinal plasma density profiles are taken into account, notably the fact that where the most marked filaments are observed (see Fig.2.d), the plasma density is much lower than in the central flat density region of the gas jet. In the experimental conditions, the characteristic times for hydrodynamics, self-focusing (SF) and plasma heating by inverse Bremsstrahlung (IB), respectively, are $\tau_s$~ $r_L/c_s$ = 316 ps (where $c_s$ = 0.11 µm/ps is the ion acoustic velocity), $\tau_{SF}$ ~ $\tau_s/(P/P_c)^{1/2}$ = 50 ps (where $P_c$ is the SF critical power [4]) [21] and $\tau_{IB}$ ~ $6cn_cT_e/(4\nu_{ei}I_L)$ = 15 ps (where $\nu_{ei}$ is the electron-ion collision frequency). These times are smaller or of the order of the laser pulse duration (400 ps), so that the LPI is strongly non-stationary and non-isothermal. Moreover, because the electron-ion collision mean-free-path $\lambda_{ei}$ = 25 µm is comparable to $r_L$, electron heat transport is non-local. Therefore, plasma temperatures were calculated in PARAX considering IB and nonlocal electron transport (NLT) [22]. The NLT model used here is based on a linearization of the Fokker-Planck equation to determine NLT coefficients in Fourier space. Corrections of the ponderomotive force due to NLT were tested to give a negligible contribution. Strong departure from the Maxwellian distribution due to laser heating (the Langdon effect) [23], associated with a reduction of the laser absorption, is also here not considered important as our results suggest that we are here in conditions of both decreased heat flux (see later as the experimental data suggest that there are strong temperature gradients at the edge of the laser pulse) and enhanced laser absorption (based on the simulations).

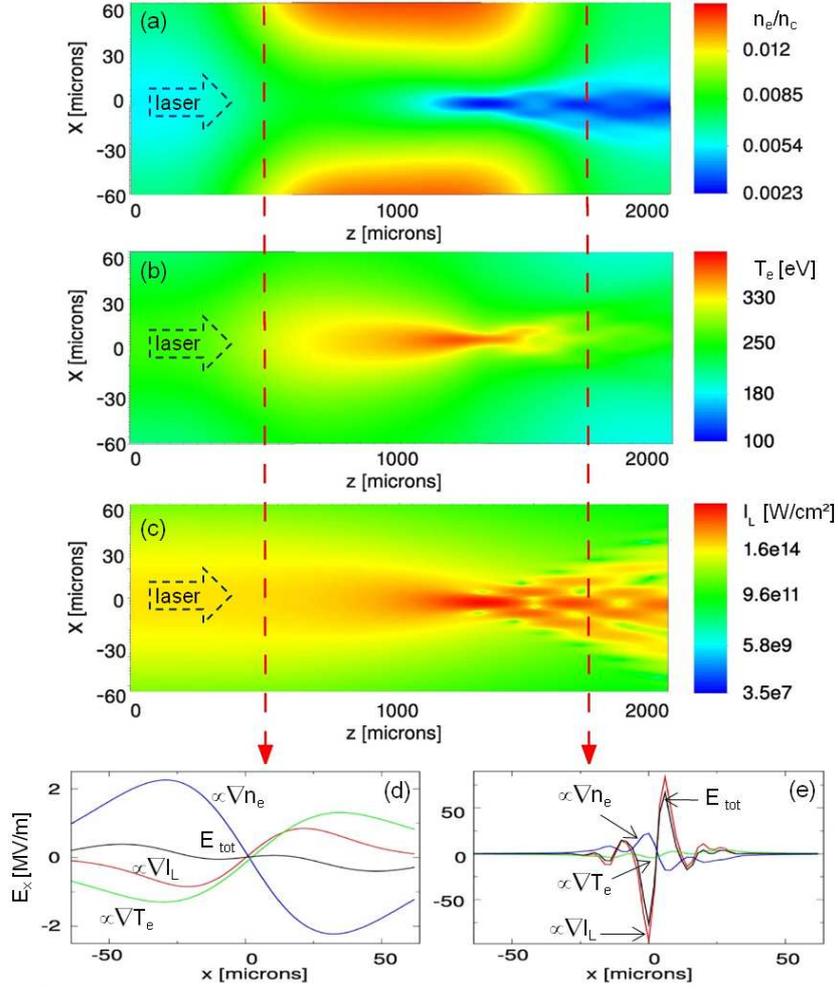

Figure 3 (color online) (a) Electron density, (b) electron temperature, and (c) laser intensity (x,z) maps at y=0, obtained from 3D PARAX simulations 150 ps after the maximum laser intensity. (d) plots of $E_x(x)$ (black line) at (y=0,z=500 µm) with ponderomotive (red), density (blue) and temperature (green) contributions. Note that the overall electron pressure contribution to the field can be obtained by adding the density and temperature components. (e) Same at (y=0,z=1700 µm).

Figure 3(a-c) show numerical results sampled at the time of Fig.2. Note however that these 2D (x,z) maps correspond to y=0 planes of the 3D laser and plasma parameters and can only be compared to Fig.2 after post-processing. In our simulations, we can observe SF after ~1200 µm, before the laser beam breaks into filaments with a characteristic size ~5 - 10 µm. This filament structure is qualitatively very similar to what is observed in Fig.2(c, d), but without the strong SF stage. This is most certainly due to plasma density fluctuations excited during plasma ionization and early heating [8], but not accounted for in PARAX as hydrodynamic codes are too dissipative to treat them correctly, that serve as a strong seed for filamentation [23].

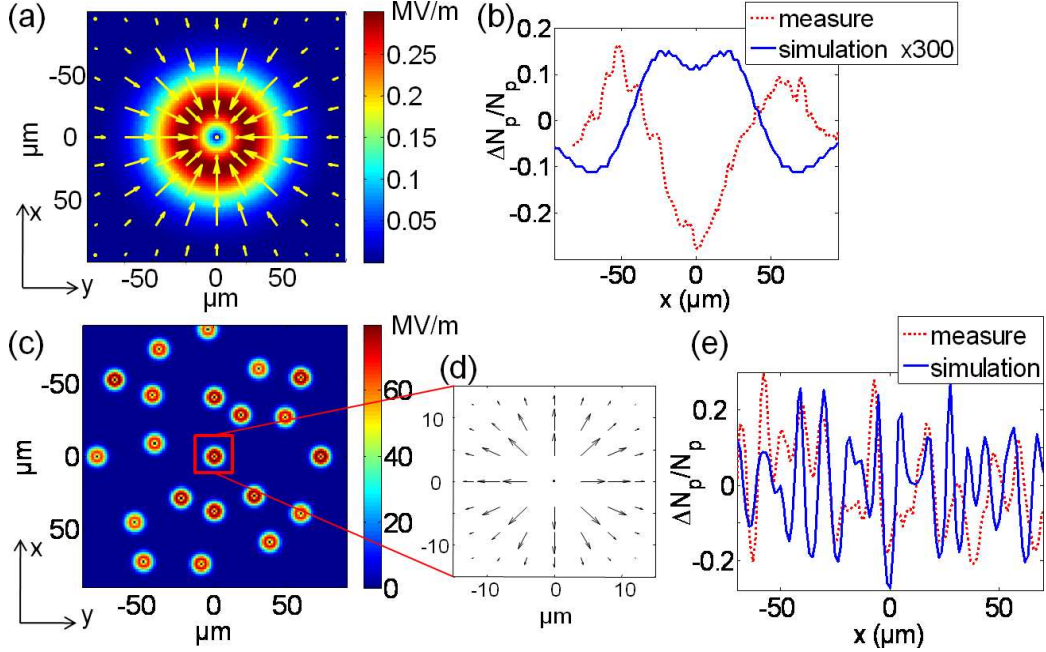

Figure 4 (Color online) (a) Map of the amplitude of the electric field $(E_x^2+E_y^2)^{1/2}$ at the beginning of propagation (z=500 µm) as retrieved from the simulations. (b) Full line: computed proton deflection through the field of (a), dashed line: measured deflections reproduced from Fig.2(b). (c) Same as (a) in the filamented region (z=1700 µm). (d) Zoom on the E-vector of one single filament. (e) Full line: computed proton deflections through the field of (c), dashed line: measured deflections reproduced from Fig.2(d).

Now to compare quantitatively the simulated plasma parameters to the measured ones, we first derive the E-fields from the simulated density, temperature, and laser intensity. Fig.4(a) shows the simulated field amplitude $(E_x^2+E_y^2)^{1/2}$ in the (x,y) plane and Fig.3(d) shows a x-lineout of just $E_x$ for y=0. Both figures are relative to the z=500 µm region, i.e. in the unperturbed region of the plasma. Fig.4(c) and Fig.3(e) show the same quantities in the filamented region (z=1700 µm). One sees that the field orientation is reversed between Fig.4(a) and Fig.4(c). This is due to the dominating $\nabla n_e$ (with $n_e$ having an on-axis dip) component of the pressure term ($E_{ep}$) for Fig.4(a) while for Fig.4(c), the $\nabla I_L$ (with $I_L$ having an on-axis peak) component of $E_x$ dominates. We also observe that the E-field strength modelled at the entrance in the plasma (~0.3 MV/m) is in strong discrepancy with the measured $E_0^{start}$.

Furthermore, we juxtapose in Fig.4(b) and (e) the simulated and measured proton deflections. The simulated deflections are obtained by tracking in 3D the protons through the 3D fields obtained from the simulations using a Monte Carlo routine and taking into account the angular spread of the proton beam. In the filamentation region (Fig.4(c)), i.e. at the end of the laser propagation in the plasma, the inhomogeneous field distribution with a maximum $E_x \sim 80$ MV/m yields a very similar pattern (Fig.4(e)) compared to what was observed experimentally in Fig.2(d). This shows that the simulation results well meet the analytical estimate for $E_x$. It is also worth noting that, in this region, thermal perturbations are smoothed (Figure 3(b)), due to heat conductivity and non-stationary filament behavior. This is further substantiated by the fact that the $\nabla n_e$ contribution to $E_{ep}$ remains small (Figure 3(e)), hence the field reduces only to the $E_{lp} \propto \nabla I_L$ component associated to each filament.

All this is in stark contrast to what is observed in the first 500 µm of propagation, i.e. before filamentation occurs. In this region, a quasi-equilibrium is built up between $E_{ep}$ and $E_{lp}$. It is however not complete due to the time variation of laser intensity. The net E-fields are therefore low, of 0.3 MV/m only (Fig.4(a)), i.e. more than two orders of magnitude smaller than $E_0^{start} \sim 110$ MV/m inferred from Fig.2(b). We already mentioned that the field is here dominated by the $\nabla n_e$ term of the pressure gradient field $E_{ep}$ (Fig.3(d)). Yet, not only is its simulated proton signal (Fig.4(b)) much weaker than observed, but also the inward directed simulated fields lead to deflections opposite to what is measured (Figure 2(b)). We therefore deduce that the simulation underestimates either the $\nabla T_e$ or $\nabla I_L$ term for the field otherwise they would lead to the correct proton deflection orientation. Furthermore, the $E_{lp}$ component itself cannot explain the observed strong field of Fig.2(b). Indeed, at the plasma entrance, $I_L$ and, thus, $E_{lp}$ are well known and cannot contribute to more than 1 MV/m (Fig.3(d)).

E-fields of ~100 MV/m have already been observed under very different conditions (very dense plasmas) [10] and were explained by very strong ionization gradients or very thin (<0.1 micron) shock fronts [24]. These collisional effects are however not good candidates to explain our observations as the He plasma is fully ionized such that ionization gradients are not expected. Moreover, current LPI conditions with a long mean free path of particles do not allow for the generation of a shock in the plasma: the slowly-varying density profile has characteristic scale lengths of tens of µm.

We have tested the hypothesis of a mechanism (e.g. involving magnetic fields) strongly inhibiting heat flow away from the axis, leaving a pressure maximum on axis, thus enhancing the resulting electrostatic fields. For this, simulations were performed with heat conductivity artificially put to zero. However electrostatic fields in these simulations were enhanced only by a factor 4, which is far from what is enough to explain the experimental results, thus suggesting that both enhanced laser absorption and reduced heat flux are necessary to explain the observed strong E-fields.

The most likely remaining explanation for the observed strong E-fields is therefore a strong $T_e$ gradient at the edge of the laser beam. The radial heat flux is sufficiently strong in our conditions for the return current to drive an ion acoustic instability [25]. The anomalous electron heating caused by this turbulence, strongly affecting the laser absorption and heat transport in the plasma, would enhance electrical resistivity and consequently result in strong E-fields [26]. Unfortunately the nonlocal electron transport under such conditions in laser plasmas has not yet been explored, because it is associated with a strong distortion of the electron and ion distribution functions that the code is unable to integrate. In summary, although the results presented here demonstrate the ability of present-day code to investigate filamentation of a high-power laser beam, they also evidence the inability of the same code to model strongly nonlinear and nonlocal heat transport. This stresses the need for development of full kinetic collisional simulations with mobile ions.

We acknowledge the expert support of the LULI teams and discussions with P. Amendt, M. Borghesi, S.N. Chen, C.K. Li and O. Willi. This work was supported by the DAAD, grant E1127 from Région Ile-de-France, and the CEA-EURATOM association as a "keep-in-touch" activity. Use of the computing center *CCRT* of the CEA and financial support from the *Agence Nationale de la Recherche* under project No. ANR-07-BLAN-004 is acknowledged. M.N. was supported by JSPS.

*Present address: INFN, Via E. Fermi, 40 - 00044 Frascati, Italy and ILE - Ecole Polytechnique - CNRS –ENSTA – Iogs - UP Sud, Batterie de l'Yvette, 91761 Palaiseau, France